%% file: main.tex
\documentclass{article}
\usepackage{spconf,amsmath,graphicx,hyperref}
\usepackage{booktabs}
\usepackage{stfloats}
\usepackage{subcaption} 

\newcommand{\projectname}{TQCodec}
\usepackage{xcolor}

\title{\projectname{}: Towards Neural Audio Codec for High-fidelity Music Streaming}
\name{%
  \begin{tabular}{@{}c@{}}
    Lixing He$^{1,2}$ \qquad
    Zhouxuan Chen$^{1}$ \qquad
    Mingshuai Liu$^{1}$ \qquad
    Xinran Sun$^{1,3}$ \qquad
    Wucheng Wang$^{1}$ \\[1ex]
    Minfu Li$^{1,4}$ \qquad
    Lingcheng Kong$^{1\ast}$\thanks{$^{\ast}$ Corresponding author: lesterkong@tencent.com} \qquad
    Weifeng Zhao$^{1}$ \qquad
    Wenjiang Zhou$^{1}$
  \end{tabular}
}

\address{
$^{1}$ Tencent Music Entertainment, 
$^{2}$ The Chinese University of Hong Kong, 
$^{3}$ Southeast University \\
$^{4}$ Tsinghua University 
}
\begin{document}
%
\maketitle
\begin{abstract}
We propose \projectname{}, a neural audio codec designed for high-bitrate, high-fidelity music streaming. Unlike existing neural codecs that primarily target ultra-low bitrates ($\leq16 kbps$), \projectname{} operates at 44.1 kHz and supports bitrates from 32 kbps to 128 kbps, aligning with the standard quality of modern music streaming platforms. The model adopts an encoder-decoder architecture based on SEANet for efficient on-device computation and introduces several enhancements: an imbalanced network design for improved quality with low overhead, SimVQ for mid-frequency detail preservation, and a phase-aware waveform loss. Additionally, we introduce a perception-driven band-wise bit allocation strategy to prioritize perceptually critical lower frequencies. Evaluations on diverse music datasets demonstrate that \projectname{} achieves superior audio quality at target bitrates, making it well-suited for high-quality audio applications.
\end{abstract}
\section{Introduction}
\label{sec:intro}

A high-fidelity audio codec with a compact representation is essential for both audio compression and generation. While traditional codecs such as Opus and MP3 have seen widespread adoption across numerous applications, the rise of deep neural networks has inspired the development of neural audio codecs (NAC). Models like SoundStream, Encodec, and DAC have already achieved performance competitive with their conventional counterparts. Although NACs bypass handcrafted feature engineering, most existing codecs target relatively low bitrates. In scenarios such as music streaming, where available bandwidth can reach 32 kbps or higher, audio quality takes precedence over compression efficiency. This motivates the need for a new neural audio codec designed for high-bitrate, high-quality applications.

In this paper, we propose \projectname{}, a neural codec that can encode 44.1 kHz music with higher bitrates from 32kbps to 128kbps, matching the standard bitrate of music streaming. In comparison, most neural code focuses on much lower bitrates (mainly 16kbps or less). Consequently, it is necessary to explore designs that can still benefit the codec at high bitrates. 
\projectname{} models the audio in the time domain, following the well-explored encoder-decoder architecture, same as DAC \cite{kumar2023high} and Encodec \cite{defossez2022high}. Differently, we redesign the codec to well suit the limitation of end-device (i.e., mobile phones) by replacing the DAC encoder decoder with SEANet \cite{tagliasacchi2020seanet}, which has a short receptive field and low computation overhead. Besides, propose few of techniques to further boost the performance, including an imbalanced pair of networks that improve the quality while preserving lightweight computation, SimVQ that preserves the mid-frequency details, and waveform loss that compensates for the phase.
Finally, we propose a perception-driven band-wise allocation for the codec, which can distribute more resources on lower frequencies that impact the listening experience most.

\projectname{} is evaluated in large-scale and diverse music datasets with different bitrates from 32kbps to 128kbps. We present both objective evaluation and visualization on the spectrogram to validate the benefits of our design. 

\section{Related Work}
Neural audio codecs typically adopt a VQ-GAN framework with an encoder-decoder structure and a quantized bottleneck, often using residual vector quantization (RVQ). Key works include SoundStream \cite{zeghidour2021soundstream}, Encodec \cite{defossez2022high} with its multi-scale STFT discriminator, and DAC \cite{kumar2023high} featuring an enhanced RVQ design. Among these, DAC remains a strong baseline despite its earlier release.

Music presents unique challenges for audio codecs with its wide-band signals and rich harmonics, demanding both clarity and high fidelity.
The recent surge in music generation research highlights the need for advanced audio codecs. While systems like \cite{team2025live} leverage SpectroStream \cite{li2025spectrostream} for music synthesis, its 16kbps bitrate often falls short in preserving musical fidelity. Alternative approaches employing neural vocoders \cite{zhu2024musichifi} can enhance quality but incur heavy computational costs, making them impractical for real-time streaming. \projectname{} bridges this gap by delivering both efficiency and high-fidelity reconstruction tailored for music applications.

On the other hand, a significant portion of codec research focuses on speech, where low bandwidth is critical and sufficient. A direct application of those codecs is for generative tasks, where a low bitrate and single codebook are desired \cite{li2024single, xin2024bigcodec}. 
Other works extend NAC by incorporating semantic knowledge \cite{liu2024semanticodec}, while it covers general audio, including music, it still focuses on ultra-low bitrates. 
Additionally, ScoreDec \cite{wu2024scoredec} and FlowDec \cite{welker2025flowdec} have been proposed to enhance the perception quality of NAC by post-filtering with a diffusion model. However, the extra module can increase the RTF a lot, making it not realistic for music streaming.

\input{method}

\input{experiment}

\section{Conclusion}
We propose \projectname{}, a neural audio codec for high-bitrate music streaming at 44.1kHz (32-128kbps). Key innovations include: (1) SEANet-based architecture for efficient on-device decoding (6.31 GMACs), (2) SimVQ for mid-frequency preservation, (3) phase-aware waveform loss, and (4) perceptually optimized bit allocation. Evaluations show superior quality over baselines in LSD/SNR metrics while maintaining computational efficiency. The codec is particularly effective for music streaming applications requiring high fidelity.

\newpage
\bibliographystyle{IEEEbib}
\bibliography{refs}

\end{document}

%% file: method.tex
\section{Method}
Our model is built on the framework of DAC \cite{kumar2023high}, following the same end-to-end pipeline as encoder-quantizer-decoder. As shown in Tab. \ref{tab:general_compare}, \projectname{} works on a much higher bit rate.

\begin{table}[h]
    \centering
    \caption{Comparison of compression approaches (SR: sample rate, BR: bit rate, CB: codebook, FR: frame rate).}
    \vspace{-1em}
    \begin{tabular}{c|c|c|c|c}
    \toprule
        Codec & SR (khz) & BR (kbps) & CB & FR \\
        \midrule
        \projectname{} & 44.1  & 32, 64, 128 & 5/ 10/ 20 & 689\\

        DAC \cite{kumar2023high} & 44.1 & 16 & 18 & 86\\

        Encodec \cite{defossez2022high} & 24/ 48 & 24 & 16 & 75 /150\\
        \bottomrule
    \end{tabular}

    \label{tab:general_compare}
    \vspace{-1em}
\end{table}
\subsection{Improved Neural Audio Codec}

\subsubsection{Encoder and decoder}
The audio codecs involve a two-stage process: encoding and decoding. In the context of music streaming, the encoding stage is typically performed once on cloud servers, where computational resources are sufficient, making its complexity manageable and not a primary bottleneck. In contrast, the decoding stage occurs on end-user devices, predominantly smartphones with varying and often limited computational capabilities. To ensure broad accessibility and a smooth user experience, the decoding process must remain within practical computational limits. 

For \projectname{}, we consider both the computation and latency, where the former is constrained by an upper bound of 10 Giga Multiply-Accumulate Operations (GMACs) per second. Unfortunately, the default configuration of DAC's decoder has approximately 365 GMACs, significantly exceeding our limits. 
One straightforward way to reduce the number of MAC operations is to decrease the model's dimensions. For example, we could reduce the decoder dimension from 1536 to 128, which effectively reduces the computation. However, we noticed that the receptive field remains unchanged at 17704 samples, mainly due to the dilation and high downsample rate of DAC. Such a large receptive field can result in huge latency, especially when the streaming window is short.

To tackle this issue, we replaced the original encoder and decoder structures with those from Encodec, which utilizes the efficient SEANet architecture \cite{tagliasacchi2020seanet} enhanced by an LSTM layer. This architecture is specifically designed for computational efficiency, low latency, and compatibility with streaming applications.
As a result, the decoder's computation is successfully reduced to approximately 6.31 GMACs, while the receptive field has been minimized to only 2410 samples.
Additionally, we found that the encoder requires about 2 GMACs by default. Since this does not affect the decoding process, we have decided to empirically increase the encoder's computation to approximately 80 GMACs.

\subsubsection{Quantizer}
Following the design of DAC, we retain its improved residual vector quantization (RVQ) framework, which achieves high codebook utilization.
However, during our listening tests, we observe that the decoded audio still contains perceptible noise—even when the scale-invariant signal-to-noise ratio (SI-SNR) is already high.
Further analysis of the band-wise reconstruction error reveals that the RVQ struggles to accurately model the mid-frequency range (3000–8000 Hz).
To address this issue, we propose replacing the standard VQ layer within the RVQ structure with a SimVQ module, as introduced in \cite{zhu2024addressing}, whose codebook is frozen and the codes are implicitly generated through a linear projection. To enable a higher bitrate, we extend it into the residual SimVQ, where each SimVQ quantizes the residual from the previous one.

\subsubsection{Loss function}
DAC has already introduced a combination of discriminators, loss functions, and corresponding weights that facilitate stable model convergence. Specifically, there is a mel-spectrogram loss for the generator that includes both the L1 loss of magnitude and the log-magnitude. For the adversarial loss, there are a multi-period discriminator and a multi-band multi-scale STFT discriminator, along with the HingeGAN \cite{lim2017geometric} adversarial loss formulation and feature loss \cite{kumar2019melgan}. Additionally, we apply the simple codebook and commitment losses, using stop-gradients from the original VQ-VAE formulation \cite{van2017neural}. 

However, the mel-spectrogram loss used for the generator has difficulty addressing phase confusion, as noted by \cite{wu2024scoredec}. While the DAC model can utilize adversarial training with a discriminator to help alleviate this problem, we have found that incorporating a waveform loss leads to even further improvements.
Regarding the weights assigned to each loss, we apply the following weightings: 15.0 for the multi-scale mel loss, 1.0 for the waveform loss, 2.0 for the feature matching loss, and 1.0 for the adversarial loss. Additionally, we use weightings of 1.0 for the codebook loss and 0.25 for the commitment loss. These weightings are consistent with recent studies \cite{kumar2023high} regarding the mel loss.

\subsection{Perception-driven Subband Modeling}
\begin{figure}[h]
    \centering
    \includegraphics[width=1\linewidth]{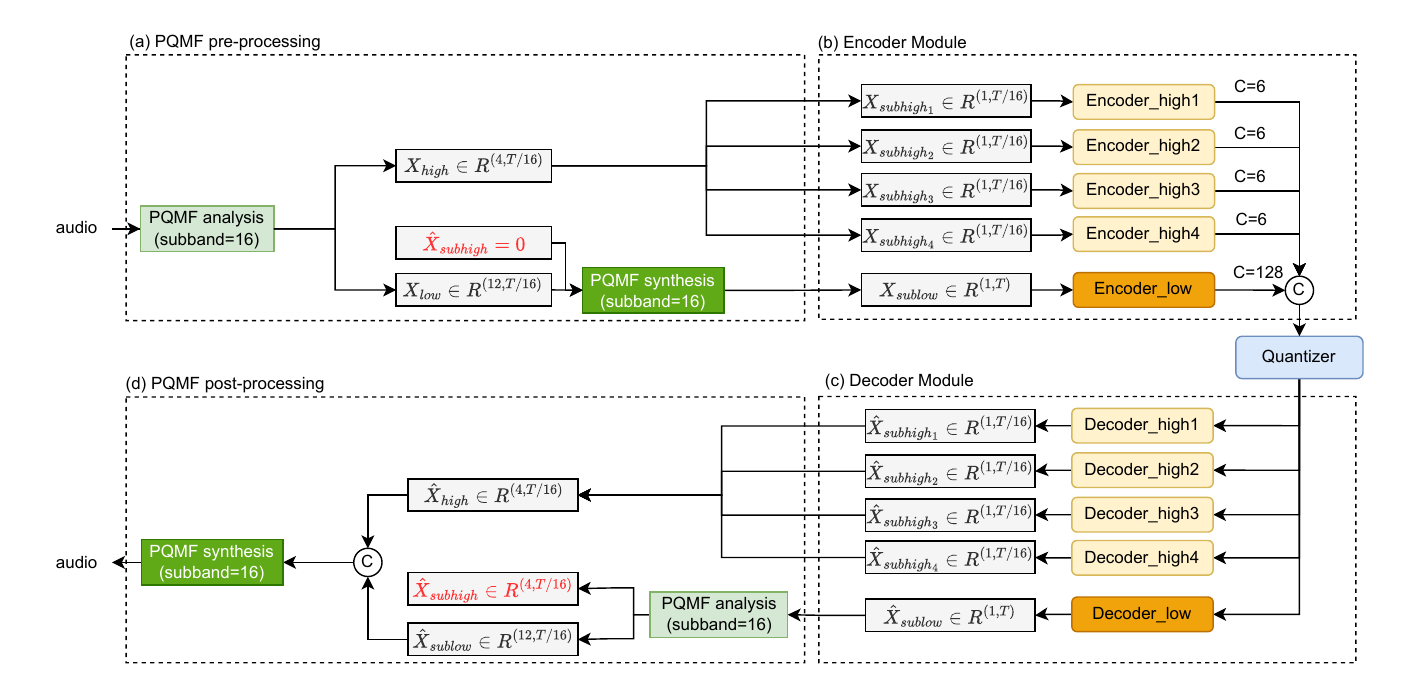}
    \vspace{-1em}
    \caption{Illustration of our subband modeling.}
    \label{fig:pqmf}
    \vspace{-1em}
\end{figure}

The human auditory system typically perceives sound frequencies ranging from 20 Hz to 20,000 Hz. However, this range is not perceived linearly—human hearing exhibits greater sensitivity to frequency differences in lower ranges than higher ones. To align more closely with human perceptual characteristics, several psychoacoustic frequency scales have been developed, including the Mel scale and the Bark scale. Taking Opus \cite{moffitt2001ogg} as an example, where the bark scale is utilized, and the algorithm will drop the high frequency when the bitrate is not sufficient. 

In the design of \projectname{}, as illustrated in Fig. \ref{fig:pqmf}, we adopt a structure similar to that of our neural audio codec. It is worth noting that the NAC operates via 1-D convolution on raw waveform data, setting it apart from spectrogram-based codecs that model each frequency band explicitly. Instead, we employ a Polyphase Quadrature Modulation Filter (PQMF) to decompose the input waveform into 16 subbands.
The first 12 subbands, designated as the core band, are perceptually most significant for human hearing and are allocated the majority of computational resources. We maintain the same network dimension for the core band as in previous configurations, while reducing the dimension of the remaining subbands to one-fourth, resulting in only an extra 0.17 MACs.
To maintain a consistent bitrate, latent representations from all subbands are concatenated prior to quantization, with the core band assigned a dimension of 128 and each of the other subbands a dimension of 6.

%% file: experiment.tex
\section{Experiment}

\subsection{Training recipe}

\begin{table}[h]
    \centering
    \caption{The dataset information.}
    \vspace{-1em}
    \begin{tabular}{ccc}
    \toprule
        Name & Number of tracks \\
        \midrule
        MusDBHQ \cite{rafii2019musdb18} & 150 \\
        Jingju \cite{gong2017creating} & 120 \\
        Jamendo \cite{bogdanov2019mtg} & 55,609 \\
        Fma \cite{defferrard2016fma} & 106,574\\
        Private dataset & 100000+ \\
        \bottomrule
    \end{tabular}
    \label{tab:dataset}
    \vspace{-1em}
\end{table}

We train and test \projectname{} with both public and private datasets, whose details are listed in Tab. \ref{tab:dataset}.
In the default experiment, we train each model with a batch size of 32 for 400k iterations on an $8 \times$ H20 machine. In practice, this takes about 30 hours to finish the training. During training, we set the duration of the audio sample to be one second. We use the AdamW optimizer with a learning rate of 0.0001, $\beta_1$ = 0.8, and $\beta_2$ = 0.9, for both the generator and the discriminator.

For the 64 kbps stereo configuration (32kbps for each channel), the model employs 3 encoder blocks and 3 decoder blocks with downsampling factors of [2,4,8]. It uses 5 codebooks, each of size 512, resulting in a bitrate of $44100/(2*4*8)*5*\log_2(512) = 31007$. For the 128 kbps stereo configuration, the number of codebooks is increased to 9, yielding a bitrate of $44100/(2*4*8)*10*\log_2(512) = 62015$. The encoder dimension is set to 64, the latent dimension to 128, and the decoder dimension to 128. Note that the actual bitrate does not exactly match the expected values but lower than it, primarily due to the 44.1 kHz sample rate.
For the two-channel audio, we apply \projectname{} twice since we empicially find that a two-channel neural network doesn't perform well. Since the loss and discriminator design are similar to DAC, we keep the same configurations of \cite{kumar2023high}.


\textbf{Evaluation metrics}:
We utilize Log-Spectral Distance (LSD) as our main metric since it focuses more on the full-band performance rather than the low-band dominated one.
Specifically, it measures frequency-domain quality between reconstructed and ground truth audio:
\(\text{LSD}(x, y) = \frac{1}{L} \sum_{l=1}^{L} \sqrt{\frac{1}{K} \sum_{k=1}^{K} \left( X(l,k) - \hat{X}(l,k) \right)^2},\)
where $l$ and $k$ are time and frequency indices, $X = \log(| \text{STFT}(y) |^2)$, and $\hat{X} = \log(| \text{STFT}(x) |^2)$.
Besides, we also utilize SNR as our metric, which measures signal quality relative to noise, computed as
\(\text{SNR}(x, y) = 20 \log_{10} \left( \left( \frac{y}{x - y} \right)^2 \right),\)
where $x$ is the estimated audio and $y$ is the clean audio. 

\begin{figure*}[ht]
    \centering
    \begin{subfigure}[b]{0.24\linewidth}
        \centering
\includegraphics[width=\textwidth]{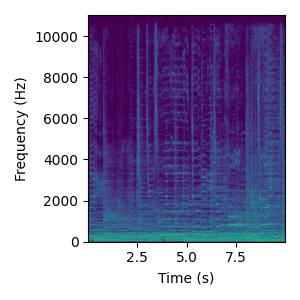}
\vspace{-1.5em}
        \caption{Clean}
        \label{fig:1a}
    \end{subfigure}
    \hfill
    \begin{subfigure}[b]{0.24\linewidth}
        \centering
    \includegraphics[width=\textwidth]{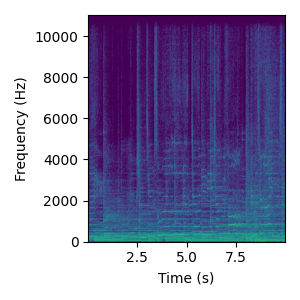}
    \vspace{-1.5em}
        \caption{RVQ}
        \label{fig:1b}
    \end{subfigure}
    \hfill
    \begin{subfigure}[b]{0.24\linewidth}
        \centering
    \includegraphics[width=\textwidth]{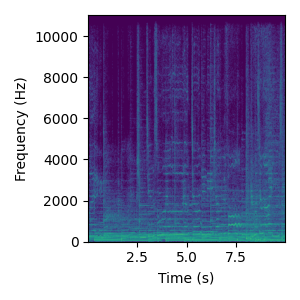}
        \vspace{-1.5em}
        \caption{RSimVQ}
        \label{fig:1c}
    \end{subfigure}
    \hfill
    \begin{subfigure}[b]{0.24\linewidth}
        \centering
\includegraphics[width=\textwidth]{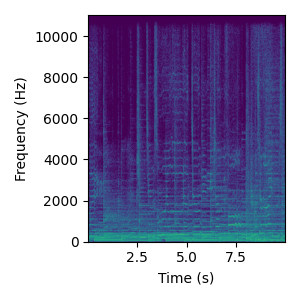}
\vspace{-1.5em}
    \caption{Subband modeling}
        \label{fig:1d}
    \end{subfigure}
    \vspace{-1em}
    \caption{Spectrogram visualization, where the RVQ fails to reconstruct the mid-frequency clearly (above 4000Hz).}
    \label{fig:MEL}
    \vspace{-1em}
\end{figure*}
\subsection{Evaluation}

\begin{table*}[h]
\centering
\caption{Ablation study results of the proposed codec, where LSD-L refers to frequency below 16000Hz and LSD-H refers to frequency above 16000Hz.}
\vspace{-1em}
\label{tab:ablation_study}
\begin{tabular}{lcccccccc}
\toprule
\textbf{Name} & \textbf{LSD $\downarrow$} & \textbf{LSD-L$\downarrow$} & \textbf{LSD-H $\downarrow$} & \textbf{MACs-E $\downarrow$} & \textbf{MACs-D $\downarrow$} & \textbf{Receptive field $\downarrow$}\\
\midrule
DAC (16kbps) & 0.927 & 0.877 & 1.016 & 25G & 365G & 17706 \\
DAC-small (32kbps) & 0.833 & 0.781 & 0.936 & 25G & 365G & 17706 \\
\midrule
\textbf{Ablation on \projectname{} (32kbps)} \\
\quad Encodec + DAC RVQ & 0.854 &  0.790 &  0.983 & 2.4G & 6.31G & 2410 \\
\quad + RSimVQ & 0.850 & 0.772 & 0.997  & 2.4G & 6.31G & 2410 \\
\quad + Waveform loss &  0.844  &  0.769 & 0.995 & 2.4G & 6.31G & 2410 \\
\quad + Inbalanced Autoencoder & \textbf{0.822} & \textbf{0.757} & \textbf{0.949} & 72G & 6.31G & 2410\\
\midrule
\quad 64kbps & 0.7715 & 0.647 & 0.950 & 72G & 6.31G & 2410\\
\quad 128kbps & 0.6716 & 0.5166 & 0.9517 & 72G  & 6.31G & 2410 \\
\bottomrule
\end{tabular}
\vspace{-1em}
\end{table*}

\projectname{} is composed of several different components, making it essential to assess the performance improvements gained from adding each one, as demonstrated in Table \ref{tab:ablation_study}. We begin by integrating the Encodec encoder and decoder with the enhanced RVQ from DAC in the first row, which yields competitive performance compared to the baseline while requiring significantly less computation. Next, we sequentially add the RSimVQ, waveform loss, and the imbalanced autoencoder. Our findings reveal that the imbalanced autoencoder is the most impactful component in terms of LSD improvement, especially for the LSD in the lower frequency range. It is important to note that none of the ablation components necessitate additional computation during the decoding process. At the end of the Table, we scale up \projectname{} to 64kbps and 128kbps, and we observe that the LSD-L keeps decreasing while the LSD-H is already saturated.

Moreover, we analyze the impact of RSimVQ on mid-frequency reconstruction. As indicated in Table \ref{tab:ablation_study}, RSimVQ effectively reduces LSD-Low. More importantly, we observe that it mitigates artifacts that adversely affect listening quality, as illustrated in Fig. \ref{fig:MEL}. Although these artifacts may not be fully reflected in objective metrics, they notably influence subjective perception.
Specifically, while the RVQ output in Fig. \ref{fig:1b} exhibits a blurred spectrogram between 5000–10000 Hz, RSimVQ successfully restores certain harmonic structures in this region, as shown in Fig. \ref{fig:1c}.

In our analysis of subband modeling, we compare \projectname{} with Ogg-Vorbis \cite{moffitt2001ogg} as presented in Tab. \ref{tab:subband}. We observed that Ogg-Vorbis significantly discards high-frequency content at a bitrate of 32kbps, leading to a much lower LSD-High score. However, by utilizing our subband modeling technique, we achieve an improvement in both LSD-L, and SNR, while only slightly increasing the LSD-High. As illustrated in Fig. \ref{fig:1d}, our subband modeling maintains a similar harmonic structure while greatly enhancing the quality of low frequencies.

\begin{table}[h]
\centering
\caption{Comparison with baselines to verify the subband modeling.}
\vspace{-1em}
\label{tab:subband}
\begin{tabular}{cccccc}
\toprule
\textbf{Model} & \textbf{Bitrates} & \textbf{LSD-L $\downarrow$} & \textbf{LSD-H $\downarrow$} & \textbf{SNR $\uparrow$}\\
\midrule
Ogg-Vorbis \cite{moffitt2001ogg} & 48kbps  & 0.763 & 2.847 & 16.998\\
\midrule
\projectname{} & 43kbps & 0.718 & 0.954 & 16.074 \\
+ Subband & 43kbps & 0.702 & 1.163 & 16.88\\
\bottomrule
\end{tabular}
\vspace{-1em}
\end{table}

\begin{table}[h!]
\centering
\caption{Subjective Listening Test: AICodec-64kbps vs. Ogg-96kbps (HQ)}
\vspace{-1em}
\begin{tabular}{|l|c|}
\hline
\multicolumn{2}{|c|}{\textbf{MOS}} \\
\hline
Average & 4.18  \\
Pass Rate & 100\% \\
Excellent Rate & 77.8\% \\
\hline
\multicolumn{2}{|c|}{\textbf{Preference}} \\
\hline
HQ (Ogg) Better & 39.4\%  \\
AICodec Better & 39.1\%   \\
No Significant Difference & 21.5\%   \\
\hline
\end{tabular}
\vspace{-1em}
\label{tab:full_eval}
\end{table}